\documentclass[Royal]{sagej}

\usepackage{moreverb,url}
\usepackage[colorlinks,bookmarksopen,bookmarksnumbered,citecolor=blue,urlcolor=blue]{hyperref}
\usepackage{tikz-opm}
\usepackage{amsmath, amsthm, amscd, amsfonts,graphicx}
\newtheorem{theorem}{Theorem}[]

\theoremstyle{definition}
\newtheorem{definition}[theorem]{Definition}

\newcommand\BibTeX{{\rmfamily B\kern-.05em \textsc{i\kern-.025em b}\kern-.08em
T\kern-.1667em\lower.7ex\hbox{E}\kern-.125emX}}

\begin{document}

%\runninghead{}

\title{Dynamic system of strategic games}
%\itshape{in the Second Persian Gulf War}}

%\itle{The Dynamic System of Interaction Between America and Iran in the Second Persian Gulf War}
\author{Madjid Eshaghi Gordji\affilnum{1} and Gholamreza Askari\affilnum{1}}

\affiliation{\affilnum{1}Department of Mathematics, Semnan University P.O. Box 35195-363, Semnan, Iran}
%\affilnum{2}Department of Civil Engineering, Semnan University P.O. Box 35195-363, Semnan, Iran}

%\corrauth{Madjid Eshaghi Gordji.}

\email{meshaghi@semnan.ac.ir; g.askari@semnan.ac.ir}

\begin{abstract}
Maybe an event can't be modeled completely through one game but there is more chance with several games. With emphasis on players' rationality, we present new properties of strategic games, which result in production of other games. Here, a new attitude to modeling will be presented in game theory as dynamic system of strategic games and its some applications such as analysis of the clash between the United States and Iran in Iraq will be provided. In this system with emphasis on players’ rationality, the relationship between strategic games and explicitly the dynamics present in interactions among players will be examined. In addition, we introduce a new game called trickery game. This game shows a good reason for the cunning of some people in everyday life. Cooperation is a hallmark of human society. In many cases, our study provides a mechanism to move towards cooperation between players.

\end{abstract}


\maketitle

\section{Introduction}
In recent years, number of international conflicts have been increased and this issue can lead to dangerous political games. Countries prefer to avoid military confrontation, this problems settled through peaceful negotiation. Cooperation is only profitable to each government if the other government reciprocates \cite{keohane1986reciprocity, kydd2000trust}. For this purpose, from game models are used to describe strategic interaction between countries \cite{lejano2012modeling, micozzi2016empirical}. There are aspects to a conflict which cannot be properly modelled using static analysis. It is necessary to use a dynamic model for games where there are constraints on players’ actions as time passes \cite{fraser1982dynamic}.

In game theory, players are two groups, first class is rational players and the second class is irrational players. The environment in which rational players interact is called strategic environment. The essential assumption of a player's rationality is that he/she considering his decision probable impact on other players, makes a decision along with his own benefits  \cite{belot2013players, brams1988game, van1984relation}. Each player thinks about the game continuation  according to his rationality. The environment in which irrational players interact is called evolutionary environment  \cite{nowak2006five, smith1973lhe}.

Study of games can be classified into two groups. First class is the studies that speak about simple games with a few players and a few possible actions and way of modeling by one game \cite{nash1951non, nash1950equilibrium, selten1975reexamination, von2007theory}. Second class examines relationship between games. Most of the researches conducted so far are of the  first type and second type research is very rare. Among the most important research of second type can be \textit{Meta games} and \textit{topology of the $2\times 2$ games}. In Meta games initially a game is selected and using Meta strategies of $n$ type, Meta games are developed and this structure is used to analyze game \cite{howard1971paradoxes}. In topology of  the $2\times 2$ games, the games are classified and using reflections, rotations and moves from one game reach to another and then the model result is applied in applied problems \cite{robinson2005topology}. At the \textit{Theory of Moves}, Steven Brams write: TOM is by no means the be-all and end-all of applied game-theoretic modeling. The dynamic analysis of ordinal games still has gaps that need to be filled and details that need to be worked out \cite{brams1994theory}.

 At the \textit{Theory of Games and Economic Behavior}, von Neumann and Morgenstern write: We repeat most emphatically that our theory is thoroughly static. A dynamic theory would unquestionably be more complete and therefore preferable \cite{von2007theory}. So, here we examine relationship between strategic $2\times 2$ games by presenting a new system called \textit{dynamic system of strategic games}. The dynamic system of strategic games is a dynamic model of $2\times 2$ games. The most important significance of this modeling is consideration of the impact of these games on each other. In this modeling, games, strategies, and the pair of rational actions that are created in the heart of a game will be raised and examined.

 As an application of dynamic system of strategic games, we seek modeling of conflict between United States of America and Islamic Republic of Iran in Iraq before attack to Iraq until complete withdrawal of U.S. troops in 2011. During Cold War, Iraq was one of the few allies of Soviet at the critical region of Middle-East. During 1980s, Baghdad-Washington relations s improved and reached a very good position. Yet, Iraq invasion of Kuwait in 1991 put it in a full hostility with America and finally after September 11 attacks, America decided to overthrow Saddam regime. The US invasion of Iraq and the policy guidelines that led to it thus reflect a world view whereby the United States is thought to be both so powerful and so benevolent that it has the ability to spread democracy throughout the world, which can be achieved by military force if necessary \cite{heinze2008new}. The US invasion of Iraq in 2003 is likely to become one of the most consequential American foreign policy decisions of our time \cite{boyle2004utopianism, heinze2008new}. After the war termination by collapse of Iraq Baathist regime, America entered into Iraq and took its control. New Iraq has found a position different from Iraq of Saddam Hossein period in foreign policy of Iran. Iran considered the circumstances is appropriate to enter in Iraq in view of being adjacent with this country, collapse of Baathist regime, help to people suffered from war and establishment of government in Iraq and entered in Iraq. Both countries also belong to different models of international relation and political systems.
 In the following, we introduce Trickery game. This game is an asymmetric game $2\times2$ which one of players can with cunning change his action that reduces the payoff other player. Game theorists have introduced a variety of games to express the circumstances of the event.

\section{{Results}}

\subsection*{\textbf{Concepts and terms. Game-maker games}} Here we examine the games dynamic system with rational players.  Before start of modeling, we present required concepts and terms. If a game produces other games, it is called \textit{a game-maker game}. In general, if the games $g_1, g_2, ...,g_n$ generate games $g^{'}_1, g^{'}_2, ...,g^{'}_m$, then $g_i$- and $g^{'}_i$-s are called \textit{producer} and \textit{produced}, respectively. We call the form of displaying game-maker games as \textit{dynamic system of strategic  games}.

\subsection*{\textbf{Strategy-maker game}} We consider strategic $2\times 2$ games with perfect information. If a game create one or more strategies is will be called \textit{strategy-maker game}. Each \textit{strategy} has at least two pairs of actions. Each pair of actions includes players' payoffs. The produced strategies can be dominant strategy, dominated strategy, weakly dominant strategy and weakly dominated strategy. Therefore, each dominant action of a player can be called \textit{dominant strategy} of a player. If a game doesn't generate any strategy, the game isn't strategy maker.

In a strategic game with ordinal preferences, player $i$s action  $a^{''}_{i}$ strictly dominates her action  $a^{'}_{i}$ if
\begin{equation*}
u_i(a^{''}_{i}, a_{-i})> u_i(a^{'}_{i}, a_{-i}) \;\  for \;\ every \;\ a_{-i}\in A_{-i},
\end{equation*}
where  $u_i$ is a payoff function that represents player i’s preferences \cite{osborne2004introduction}. If for player $i$ the action $a^{''}_{i}$ is preferred to action $a^{'}_{i}$ per every choice of action of other players, it is called \textit{dominant strategy} and is shown by $S^{j}_{i}$ where $S^{j}_{i}$ shows $j$-th strategy of $i$-th player.

In a strategic game with ordinal preferences, player $i$’s action $a^{''}_{i}$ weakly dominates her action $a^{'}_{i}$ if
\begin{equation*}
u_i(a^{''}_{i}, a_{-i})\geq u_i(a^{'}_{i}, a_{-i}) \;\  for\; every \;\  a_{-i}\in A_{-i}
\end{equation*}
and
\begin{equation*}
u_i(a^{''}_{i}, a_{-i})> u_i(a^{'}_{i}, a_{-i}) \;\ for \;\ some \;\ a_{-i}\in A_{-i},
\end{equation*}
where  $u_i$ is a payoff function that represents player i’s preferences \cite{osborne2004introduction}. If for player $i$ the action $a^{''}_{i}$ is preferred over action $a^{'}_{i}$ for each action choice of other players, it is called \textit{weakly dominant strategy} and will be represented by $S^{j}_{i}$.
\begin{figure}
\centering
\begin{tikzpicture}
\node [ opmobject] (4){\begin{tabular}{c|c|c|}
\multicolumn{1}{c}{$g_1$} & \multicolumn{1}{c}{} & \multicolumn{1}{c}{} \\[-2.5mm]
  \multicolumn{1}{c}{} & \multicolumn{1}{c}{C} & \multicolumn{1}{c}{D} \\ \cline{2-3}
   C & 3,2 & 2,4 \\ \cline{2-3}
   D & 4,1 & 1,3 \\ \cline{2-3}
\end{tabular}};
\node [ opmobject, right=of 4, xshift=4 pt, ] () {\begin{tabular}{c|c|c|}
 \multicolumn{1}{c}{$g_2$} & \multicolumn{1}{c}{} & \multicolumn{1}{c}{} \\[-2.5mm]
  \multicolumn{1}{c}{} & \multicolumn{1}{c}{C} & \multicolumn{1}{c}{D} \\ \cline{2-3}
   C & 1,-1 & -1,1 \\ \cline{2-3}
   D & -1,1 & 1,-1 \\ \cline{2-3}
\end{tabular}};
\end{tikzpicture}
\caption{Strategy maker games. The Bully game $g_1$ is a strategy maker game of order $(2, 1)$.  Matching Pennies $g_2$ is a strategy maker game of order $(2 , 0)$.}
\label{fig:frod}
\end{figure}

 If a game with $n$ players is strategy maker for $k$ players ($1\leq k \leq n$) it is called \textit{strategy maker game of order} $(n, k)$. If a game with $n$ players isn't strategy maker it is called \textit{strategy maker game of order} $(n, 0)$. In other word, we can consider a strategy maker game of order $(n, 0)$ as a game which is not strategy maker. In Fig. \ref{fig:frod}, we consider row player as player $1$ and column player as player $2$. The Bully game $g_1$ is a strategy maker game of order $(2, 1)$. This game generates dominant strategy of defect $D$ and dominated strategy of cooperation $C$ for player $2$, but this game isn't a strategy maker for player $1$. Matching Pennies $g_2$ isn't a strategy maker game. The game $g_2$ is a strategy maker game of order $(2, 0)$.

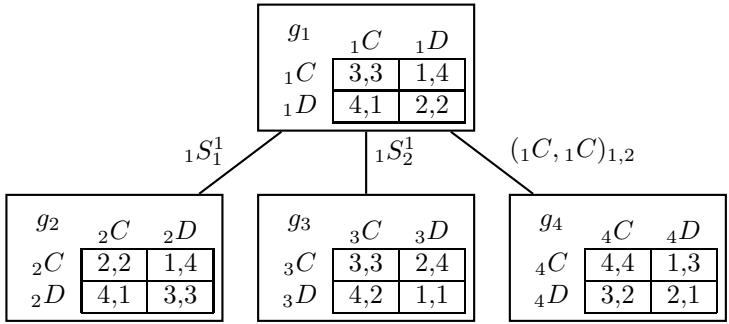
\begin{figure}
\centering
\begin{tikzpicture}
\node [ opmobject, xshift=-20] (1) {\label{tabel:label}\begin{tabular}{c|c|c|}
\multicolumn{1}{c}{$g_1$} & \multicolumn{1}{c}{} & \multicolumn{1}{c}{} \\[-2.5mm]
  \multicolumn{1}{c}{} & \multicolumn{1}{c}{${_{1}C}$} & \multicolumn{1}{c}{${_{1}D}$} \\ \cline{2-3}
   ${_{1}C}$ & 3,3 & 1,4 \\ \cline{2-3}
   ${_{1}D}$ & 4,1 & 2,2 \\ \cline{2-3}
\end{tabular}};
\node [  below=of 1, xshift=-61 pt, yshift=30 pt] (W){${ _{1}S^{1}_{1}}$};
\node [ opmobject, below=of 1, xshift=-95 pt, yshift=5 pt] (2) {\begin{tabular}{c|c|c|}
\multicolumn{1}{c}{$g_2$} & \multicolumn{1}{c}{} & \multicolumn{1}{c}{} \\[-2.5mm]
  \multicolumn{1}{c}{} & \multicolumn{1}{c}{${_{2}C}$} & \multicolumn{1}{c}{${_{2}D}$} \\ \cline{2-3}
   ${_{2}C}$ & 2,2 & 1,4 \\ \cline{2-3}
   ${_{2}D}$ & 4,1 & 3,3 \\ \cline{2-3}
\end{tabular}};

\node [  below=of 1, xshift=11 pt, yshift=30 pt] (Q){${ _{1}S^{1}_{2}}$};
\node [ opmobject, below=of 1, xshift=0 pt, yshift=5 pt] (3) {\begin{tabular}{c|c|c|}
\multicolumn{1}{c}{$g_3$} & \multicolumn{1}{c}{} & \multicolumn{1}{c}{} \\[-2.5mm]
  \multicolumn{1}{c}{} & \multicolumn{1}{c}{${_{3}C}$} & \multicolumn{1}{c}{${_{3}D}$} \\ \cline{2-3}
   ${_{3}C}$ & 3,3 & 2,4 \\ \cline{2-3}
   ${_{3}D}$ & 4,2 & 1,1 \\ \cline{2-3}
\end{tabular}};
\node [  below=of 1, xshift=78 pt, yshift=30 pt] (Q){$({_{1}C}, {_{1}C})_{1,2}$};
\node [ opmobject, below=of 1, xshift=95 pt, yshift=5 pt] (4) {\begin{tabular}{c|c|c|}
\multicolumn{1}{c}{$g_4$} & \multicolumn{1}{c}{} & \multicolumn{1}{c}{} \\[-2.5mm]
  \multicolumn{1}{c}{} & \multicolumn{1}{c}{${_{4}C}$} & \multicolumn{1}{c}{${_{4}D}$} \\ \cline{2-3}
   ${_{4}C}$ & 4,4 & 1,3 \\ \cline{2-3}
   ${_{4}D}$ & 3,2 & 2,1 \\ \cline{2-3}
\end{tabular}};
\path  (2) edge  (1);
\path  (3) edge (1);
\path  (4) edge (1);
\end{tikzpicture}
\caption{Game-maker game. Each player has two action ${_{k}C}$ or action ${_{k}D}$, where $({_{k}C}, {_{k}D})_{i,j}$ shows rational actions pair of players $i$ and $j$ from $k$-th game and ${ _{k}S^{j}_{i}}$ shows $j$-th strategy of player $i$ from $k$-th game. }
\label{fig:fros}
\end{figure}

The following example shows how a strategy maker game can produce other games. Consider two players that play the Prisoner's Dilemma $g_1$ in Fig. \ref{fig:fros}. Each player has two actions. Players can choose cooperat action ${_{1}C}$ or defect action ${_{1}D}$, where ${_{k}C}$ and ${_{k}D}$ shows actions of players from $k$-th game. Players choosing each action obtain a payoff. In the game $g_1$ for player $1$, the dominant strategy ${ _{1}S^{1}_{1}}$ is defect and dominated strategy ${ _{1}S^{2}_{1}}$ is cooperation. In this game, for player $2$ the dominant strategy ${ _{1}S^{1}_{2}}$ is defect and dominated strategy ${ _{1}S^{2}_{2}}$ is cooperation. In other words, the Prisoner's Dilemma is a strategy maker game of order $(2, 2)$. Player $1$ can do the game continuation process by strategy ${ _{1}S^{1}_{1}}$  or strategy ${ _{1}S^{2}_{1}}$. Player $2$ can do the game continuation process by strategy ${ _{1}S^{1}_{2}}$ or strategy ${ _{1}S^{2}_{2}}$. Based on the assumption of players rationality, player $1$ selects strategy ${ _{1}S^{1}_{1}}$ and player $2$ selects strategy ${ _{1}S^{1}_{2}}$ to continue the game. The strategy ${ _{1}S^{1}_{1}}$ ends to deadlock $g_2$, i.e player $1$ has designed game $g_2$ for game continuation. The strategy ${ _{1}S^{1}_{2}}$ ends to chicken, i.e. player $2$ has designed game $g_3$ to continue the game.

\subsection*{\textbf{Pair of rational actions}} Here we introduce pair of rational actions. Players’ preferences on pairs of rational actions are based on payoffs that they obtain.
\begin{definition}(\textit{Pair of rational actions}) A pair of actions is called \textit{rational} if at least hold true in one of the following conditions:
\begin{itemize}
\item  would be Nash equilibrium;
\item  pair of actions, would be Pareto dominant for both players over other pairs of actions;
\item  for each game that is strategy maker of order $(2, 1)$, pairs of rational actions for one player is responses to dominant strategy or weakly dominant strategy produced for other player.
\end{itemize}
\end{definition}

In a strategy maker game of order $(2, 2)$ where both players have dominant strategy and the game hasn't Pareto action pairs over Nash equilibrium, the Nash equilibrium of game is the only rational actions pair. For example, in game $g_2$ in Fig. \ref{fig:fros}, the only rational actions pair is $({_{2}D}, {_{2}D})$.

In Fig. \ref{fig:fros}, chicken $g_3$ is a strategy maker of order $(2,0)$. Pairs of actions $({_{3}C}, {_{3}C})$, $({_{3}C}, {_{3}D})$ and $({_{3}D}, {_{3}C})$ are rational for players. In Fig. \ref{fig:fros}, the Low Conflict game $g_4$ is a strategy maker of order $(2,1)$. Dominant strategy of player $2$ is cooperation ${_{4}C}$. Player $1$ response to this strategy makes pairs of his rational actions. Therefore, pairs of actions $({_{4}C}, {_{4}C})$ and $({_{4}D}, {_{4}C})$ are rational for player $1$. So player $1$ can select one of the pairs of rational actions according to his rationality to continue the game.  In Fig. \ref{fig:frod}, Matching Pennies $g_2$, the strategy maker is of order $(2,0)$ and also there isn't Pareto pair of actions property, on the other hand there isn't Nash equilibrium.  Therefore pair of the actions haven't rationality.

Classic dynamic games are ones in which players make decision consecutively, that is, each player must make his choice after previous player's choice. The extensive form is applied to show a classic dynamic game \cite{cressman2014replicator, harsanyi1986rational, huang2012emergence}.  As mentioned above, game-maker games display form is called dynamic system of games. Every dynamic system of games includes players set, strategies set, set of rational actions pairs, system history, node preference and systemic preference of players.

A tools that can display dynamic system of strategic games is games graph. Within each node of a graph, there is a strategic game in which players can make decisions. Each node  of this graph can be generator of the next game through the two following methods and be connected to it:
\begin{enumerate}
\item  strategies,
\item  pair of rational actions.
\end{enumerate}

Players to move from one node to another nod proceed by selecting strategy or pair of rational actions. Moving to the next node by the made strategy is always preference of one of players but continuing game with rational action pair may be preference of one player or both of them. In Fig. \ref{fig:fros}, game $g_1$ through strategies is connected to games $g_2$ and $g_3$ and through pair of rational actions is connected to games $g_4$. Strategy ${ _{1}S^{1}_{1}}$ is preference of player $1$ and strategy ${ _{1}S^{1}_{2}}$ is preference of player $2$. Pair of rational actions $({_{1}C}, {_{1}C})$ is preference of two players. In fact, methods $(1)$ and $(2)$ are edges of the concerned graph and each edge is created by decision and preference of one or two players.

In each node, player can decide whether to move to the next node or not. Nodes that a player has built through strategy or pair of rational actions, desire to continue the game are called \textit{move node}. A number of edges originate from each move node. Each edge may end to a move node or a \textit{final node}. Final node is a node that players have no appetence to continue. If an edge ended to a final node, play (system) finishes in that edge. If an edge isn't end to a final node, games system continues yet. Also it is possible that some edges end to one node. In the case that all edges end to final nodes, games system finishes.

Now, we introduce a model of dynamic system of strategic games. Also definitions and components of system is stated formally.

%%%%%%%%%%%%%%%%%%%%%%%%%%%%%%%%%%%%%%%%%%%%%%%%%%%%%%%%%%%%%%%%%%%%%%%%%%%%%%%%%%%%%%%%%%%%%%%%%%%%%%%%%%%%%%%%%%%%%%%%%%%%%%%%%%%%%%%%%%%%%%%%%%%%%%
%%%%%%%%%%%%%%%%%%%%%%%%%%%%%%%%%%%%%%%%%%%%%%%%%%%%%%%%%%%%%%%%%%%%%%%%%%%%%%%%%%%%%%%%%%%%%%%%%%%%%%%%%%%%%%%%%%%%%%%%%%%%%%%%%%%%%%%%%%%%%%%%%%%%%%

\section{Description of the model}
Dynamic system of strategic games is a model to examine interaction between decision makers more exactly. Every decision maker is a player in this model. To describe this system, we use the graph defined in above. In each node of graph there is  a strategic game with perfect information. A strategic game includes players set, actions set and preferences on pairs of actions.

Now using graphs, we introduce mathematical model as follows. Graph $\mathcal{G}$ is binary of $(G, M)$ that first coordinate $G=\{g_1, g_2, ..., g_n\}$ is a finite set of nodes that each node of this graph is a strategic game. Second coordinate is a finite set named edges that edges of this graph are produced strategies or pair of rational actions.

Set of all strategies produced by $k$-th game is represented by ${_{k}\mathcal{S}}={_{k}\mathcal{S}_1}\cup {_{k}\mathcal{S}_2} \cup \emptyset$. Set of all pairs of actions Players' in $k$-th game is shown with ${_{k}\mathcal{A}}={_{k}A_1}\times {_{k}A_2}$. Set of all pairs of rational actions  for player $i$ is shown with ${_{k}\mathcal{A}^{'}_{i}}$ that is a subset of  ${_{k}\mathcal{A}}$, for all $k\in\{1,2,...,n\}$.

Let $\mathcal{A}={_{1}\mathcal{A}} \cup {_{2}\mathcal{A}} \cup ... \cup {_{n}\mathcal{A}}\cup \emptyset$ and $\mathcal{S}={_{1}\mathcal{S}}\cup {_{2}\mathcal{S}} \cup...\cup {_{n}\mathcal{S}}$ be two sets. The set valued functions, rational actions pair $\phi^{'}_{i}: G\to \mathcal{A}$ and strategy maker $\phi_i: G\to \mathcal{S}$ for players $i$'s are defined as follows:
\begin{equation*}
\phi^{'}_{i}(g_k)={_{k}\mathcal{A}^{'}_{i}}=
\left\{\begin{array}{rl}
\{({_{k}a_i}, {_{k}a_{-i}})_i | ({_{k}a_i}, {_{k}a_{-i}})_i\in {_{k}\mathcal{A}}\} \ \ \ \ if~g_k~has\\
pair~of~rational~actions\\
\emptyset \ \ \ \ ~if~g_k~has'nt~pair~of~rational~actions,
\end{array}\right.
\end{equation*}

\begin{equation*}
\phi_{i}(g_k)={_{k}\mathcal{S}_i}=
\left\{\begin{array}{rl}
\{{_{k}S^{j}_{i}}| {_{k}S^{j}_{i}}\in {_{k}\mathcal{S}}\} \ \ \ \ \ \ if~g_k~is~strategy~maker\\
for~player~i \\
\emptyset \ \ ~if~g_k~is'nt~strategy~maker~for~player~i,
\end{array}\right.
\end{equation*}
for all $i\in N$ and $j,k\in I=\{1,2,...,n\}$, where $g_k$ shows $k$-th game, $({_{k}a_i}, {_{k}a_{-i}})_i$ shows rational actions pair of $i$-th player from $k$-th game and  ${ _{k}S^{j}_{i}}$ shows $j$-th strategy of $i$-th player from $k$-th game.

Every move of system as a member of set $M$ is as follows:
\begin{align*}
&M:=\Big\{m^{j}_{k}| \;\ m^{j}_{k}={_{k}S^{j}_{i}}\;\  or \;\  m^{j}_{k}=({_{k}a_i}, {_{k}a_{-i}})_i \ \ or \\
 & \ \ \ \ \ \ \ \ \ \ \ \ \ \ \   m^{j}_{k}=({_{k}a_i}, {_{k}a_{-i}})_{i,j} \ \ \forall \;\ {_{k}S^{j}_{i}}\in {_{k}\mathcal{S}_i}, \\
 & \ \ \ \ \ \ \ \ \ \ \ \ \ \ \  ({_{k}a_i}, {_{k}a_{-i}})_{i} \in {_{k}\mathcal{A}_i}, \ \  ({_{k}a_i}, {_{k}a_{-i}})_{j} \in {_{k}\mathcal{A}_j} \Big\},
\end{align*}
where $m^{j}_{k}$ shows $j$-th move  of $k$-th game and $({_{k}a_i}, {_{k}a_{-i}})_{i,j}$ shows the pair of rational action selected by players $i$ and $j$ of $k$-th game. Players’ move function $\varphi_{i}: M \to G^2$ and $\varphi_{i,j}: M \to G^2 \cup \emptyset$ with $ \varphi_{i,j}({_{k}S^{j}_{i}})=\emptyset$ is defined as following:
\begin{equation*}
\varphi_{i}(m^{j}_{k})=
\left\{\begin{array}{rl}
 (g_k,g_p)=g_kg_p & ~~if~~ m^{j}_{k}={_{k}S^{j}_{i}} \\
 (g_k,g_q)=g_kg_q & ~~if~~ m^{j}_{k}=({_{k}a_i}, {_{k}a_{-i}})_i,
\end{array}\right.
\end{equation*}
 \begin{equation*}
\varphi_{i,j}(m^{j}_{k})=
\left\{\begin{array}{rl}
\emptyset \ \ \ \ \ \ \ \ \ \ \ \ \ \ \ & ~~if ~~~~ m^{j}_{k}={_{k}S^{j}_{i}}\\
(g_k,g_s)=g_kg_s & ~~if ~~~~ m^{j}_{k}=({_{k}a_i}, {_{k}a_{-i}})_{i,j}.
\end{array}\right.
\end{equation*}

The above function shows by what move two play nodes have been connected to each other by one or both players. Consequently, it can be said that in move $m^{j}_{k}={_{k}S^{j}_{i}}$,  nodes $g_k$ and $g_p$ have been connected through the strategy selected by player $i$ to each other. In move $m^{j}_{k}=({_{k}a_i}, {_{k}a_{-i}})_i$ the nodes $g_k$ and $g_q$ have been connected by pair of rational action selected by player $i$ to each other. In move $m^{j}_{k}=({_{k}a_i}, {_{k}a_{-i}})_{i,j}$ the nodes $g_k$ and $g_s$ have been connected through a pair of rational actions selected by players $i$ and $j$ to each other.

Consider that $H$ is a set including all series (finite and infinite) that hold true in the following conditions:
\begin{enumerate}
\item $\emptyset$ is member of $H$.
\item Sequence $\big\{m^{j}_{i},\{g_k, m^{j}_{k}\}\big\}_{i,j,k\in I }$ for all $i, j, k\in \{1,2,...,n\}$, is a member of $H$. Each member of $H$ is called a history and is represented by $h$.
\item History $h=\big\{m^{j}_{i},\{g_k, m^{j}_{k}\}\big\}_{i,j,k\in I }$ is called final history if it is infinite or there isn't $g_{k+1}$ that is a member of h.
\end{enumerate}

The set $H$ is called \textit{system history}. In Fig. \ref{fig:fros}, the system history is as follows:
\begin{align*}
&H=\Big\{\emptyset, \big\{g_1, {_{1}S^{1}_{1}}, {_{1}S^{1}_{2}}, ({_{1}C},{_{1}C})_{1,2} \big\},  \big\{{_{1}S^{1}_{1}},\{g_2\}\big\},\\ & \ \ \ \ \ \ \ \ \ \ \ \big\{{_{1}S^{1}_{2}},\{g_3\}\big\}, \big\{({_{1}C},{_{1}C})_{1,2},  \{g_4\}\big\}\Big\}.
\end{align*}

Preferences of each node of a games system that are exactly the same preferences on the pairs of a strategic game actions are called \textit{node preferences} or \textit{tactical preferences}. Preferences on strategies set or set of rational actions pair of a game is called \textit{systemic preferences} or \textit{strategic preferences}.

\begin{definition} (\textit{Dynamic system of strategic games}) A dynamic system of strategic games with perfect information including:
\begin{itemize}
 \item  a set of players
\item for each player, a set of strategies
\item for each player, a set of rational actions pair
\item system history
\item node preferences (tactical preferences) on set of all actions pairs
\item systemic preferences (strategic preferences) on strategies or pairs of rational actions.
\end{itemize}
\end{definition}

In dynamic system of strategic games, players using conditions of producer game and generated strategies and pairs of rational actions decide what move the do along with their benefits and what game they design and where they stand. Also this system allows players to select among strategies and rational actions pair which result in his most benefits based on their abilities and future conditions using available information, according to their rationality and strategic preferences. Hence, players can agree with each other on the next move and choose a move that favors all or choose a move according to personal benefits. Players can choose several moves at the same time that may one has a personal benefit and other has a collective benefit.

In Fig. \ref{fig:fros}, by starting the first round of negotiations and choosing tactical preferences in the first node, game $g_1$ or first round of negotiations ends. Game $g_1$ is strategy maker of order $(2, 2)$, that is, producer of dominant strategies ${_{1}S^{1}_{i}}$ and dominated strategies ${_{1}S^{2}_{i}}$ for playe$i$ and has pair of rational actions $({_{1}D},{_{1}D})_{1,2}$ that is game Nash equilibrium and pair of actions $({_{1}C}, {_{1}C})_{1,2}$ that is dominant Pareto compared to game Nash equilibrium for both players. In other words, game $g_1$ provide players with the above information. Based on players being rational, the player $i$ to continue game can choose his dominant strategy or pair of rational actions, Pareto dominant. Players using analysis of game $g_1$ in the first round of negotiations and information obtained from this stage determine their systemic preferences and predict their motion path. Player $1$ according to dominant strategy ${_{1}S^{1}_{1}}$ design deadlock game. Player 2 according to dominant strategy ${_{1}S^{1}_{2}}$ design chicken game. Players agreeing to choose a pair of rational action $({_{1}C}, {_{1}C})_{1,2}$ enter in win-win game and wish this process occur in the next round. Therefore, each player has designed two games to continue negotiation process in the next stage. Player $i$ uses strategy ${_{1}S^{1}_{i}}$ as a believable or unbelievable threat to continuation of the negotiation process and in the case of not reaching result  and leaving negotiations, they  will choose the strategy. Choosing systemic preference of game $g_1$ and moving from this node, players enter in the next nodes and determine their tactical preferences in the new node. Players along with their benefits in nodes $g_2$, $g_3$ and $g_4$ prefer an action that has more income and as much as possible would be along strategic benefits and preferences. This process of  choices is performed in the next stages nodes as well.

In general, most of the games existing in nature or among mans and human communities can be modeled by the dynamic system of games. For example, games dynamic system within uterus that includes games inside uterus before a baby birth or period of a person’s lifetime with all events can be considered as games
dynamic system. As another example, the diplomatic relationships between two countries during a certain period of time can be modeled by games dynamic system. Inside a dynamic system by specifying times and subjects, there is statics as well. Therefore, the existence world is a combination of both of them.

In the following example, there are conditions which show how players can with help of dynamic system of strategic games reach the satisfactory conditions for cooperation by bargaining from the conditions where there is unwanted unfair situation and dissatisfaction. For more accurate expression, the row player is called player $1$ and column player is called player $2$. In the conditions of example, players can be two countries, wife and husband, two companies, … and set of players’ actions include cooperation action ${_{i}C}$ and non-cooperation action ${_{i}D}$.

\begin{figure}
\centering
\begin{tikzpicture}
\node [ opmobject, xshift=-20] (1) {\label{tabel:label}\begin{tabular}{c|c|c|}
\multicolumn{1}{c}{$g_1$} & \multicolumn{1}{c}{} & \multicolumn{1}{c}{} \\[-2.5mm]
  \multicolumn{1}{c}{} & \multicolumn{1}{c}{${_{1}C}$} & \multicolumn{1}{c}{${_{1}D}$} \\ \cline{2-3}
   ${_{1}C}$ & 3,1 & 2,4 \\ \cline{2-3}
   ${_{1}D}$ & 4,2 & 1,3 \\ \cline{2-3}
\end{tabular}};
\node [  below=of 1, xshift=-70 pt, yshift=27 pt] (Q){(${_{1}C},{_{1}D})_{1}$};
\node [ opmobject, below=of 1, xshift=-70 pt, yshift=2 pt] (2) {\begin{tabular}{c|c|c|}
\multicolumn{1}{c}{$g_2$} & \multicolumn{1}{c}{} & \multicolumn{1}{c}{} \\[-2.5mm]
  \multicolumn{1}{c}{} & \multicolumn{1}{c}{${_{2}C}$} & \multicolumn{1}{c}{${_{2}D}$} \\ \cline{2-3}
   ${_{2}C}$ & 4,4 & 3,1 \\ \cline{2-3}
   ${_{2}D}$ & 1,3 & 2,2 \\ \cline{2-3}
\end{tabular}};
\node [  below=of 1, xshift=50 pt, yshift=27 pt] (Q){$_{1}S^{1}_{2}$};
\node [ opmobject, below=of 1, xshift=80 pt, yshift=2 pt] (3) {\begin{tabular}{c|c|c|}
\multicolumn{1}{c}{$g_3$} & \multicolumn{1}{c}{} & \multicolumn{1}{c}{} \\[-2.5mm]
  \multicolumn{1}{c}{} & \multicolumn{1}{c}{${_{3}C}$} & \multicolumn{1}{c}{${_{3}D}$} \\ \cline{2-3}
   ${_{3}C}$ & 3,3 & 2,4 \\ \cline{2-3}
   ${_{3}D}$ & 4,1 & 1,2 \\ \cline{2-3}
\end{tabular}};

\node [  below=of 2, xshift=-15 pt, yshift=25 pt] (Q){$_{2}S^{1}_{1}$};
\node [ opmobject, below=of 2,  xshift=0 pt, yshift=2 pt] (4) {\begin{tabular}{c|c|c|}
 \multicolumn{1}{c}{$g_4$} & \multicolumn{1}{c}{} & \multicolumn{1}{c}{} \\[-2.5mm]
  \multicolumn{1}{c}{} & \multicolumn{1}{c}{${_{4}C}$} & \multicolumn{1}{c}{${_{4}D}$} \\ \cline{2-3}
   ${_{4}C}$ & 4,4 & 3,2 \\ \cline{2-3}
   ${_{4}D}$ & 3,2 & 1,1 \\ \cline{2-3}
\end{tabular}};
\node [  below=of 1, xshift=10 pt, yshift=-45 pt] (Q){(${_{2}C},{_{2}C})_{2}$};
\node [  below=of 3, xshift=12 pt, yshift=25 pt] (Q){$_{3}S^{1}_{2}$};
\node [ opmobject, below=of 3,  xshift=0 pt, yshift=2 pt] (5) {\begin{tabular}{c|c|c|}
\multicolumn{1}{c}{$g_5$} & \multicolumn{1}{c}{} & \multicolumn{1}{c}{} \\[-2.5mm]
  \multicolumn{1}{c}{} & \multicolumn{1}{c}{${_{5}C}$} & \multicolumn{1}{c}{${_{5}D}$} \\ \cline{2-3}
   ${_{5}C}$ & 4,4 & 1,3 \\ \cline{2-3}
   ${_{5}D}$ & 3,1 & 2,2 \\ \cline{2-3}
\end{tabular}};
\node [  below=of 4, left=-30 pt, yshift=-15 pt] (Q){$({_{4}C},{_{4}C})_{1,2}$};
\node [  below=of 5, right=-30 pt, yshift=-15 pt] (Q){(${_{5}C},{_{5}C})_{1,2}$};
\node [ opmobject, below=of 1,  xshift=0 pt, yshift=-168 pt] (6) {\begin{tabular}{c|c|c|}
\multicolumn{1}{c}{$g_6$} & \multicolumn{1}{c}{} & \multicolumn{1}{c}{} \\[-2.5mm]
  \multicolumn{1}{c}{} & \multicolumn{1}{c}{${_{6}C}$} & \multicolumn{1}{c}{${_{6}D}$} \\ \cline{2-3}
   ${_{6}C}$ & 4,4 & 2,3 \\ \cline{2-3}
  ${_{6}D}$ & 3,2 & 1,1 \\ \cline{2-3}
\end{tabular}};
\path  (1) edge  (2);
\path  (1) edge (3);
\path  (2) edge (4);
%\path  (3) edge (4);
\path  (2) edge (5);
\path  (3) edge (5);
\path  (4) edge (6);
\path  (5) edge (6);
\end{tikzpicture}
\caption{Dynamic system of games which how players can reach the satisfactory
conditions for cooperation}
\label{fig:Peace}
\end{figure}
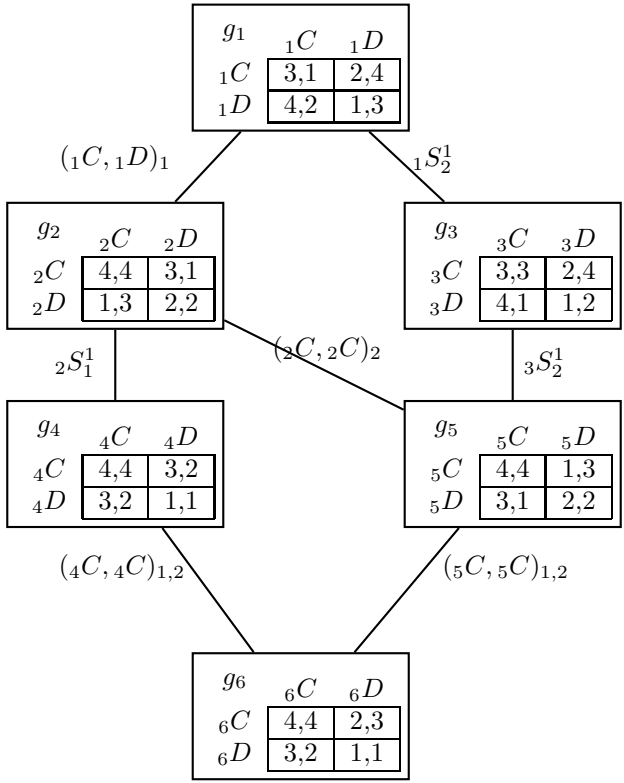

Consider two players that play Unfair game $g_1$ in Fig. \ref{fig:Peace}. Each player has two actions. Players can choose cooperation action ${_{1}C}$ or defect ${_{1}D}$. Players by choosing each action obtain a payoff. In game $g_1$ dominant strategy $_{1}S^{1}_{2}$ for player $2$ is defect and dominated  strategy $_{1}S^{2}_{2}$ is cooperation. Game $g_1$ isn't strategy maker for player $1$. In e other words, game $g_1$ is strategy maker of order $(2, 1)$. Player $2$ can continue the process of game by strategy $_{1}S^{1}_{2}$ or $_{1}S^{2}_{2}$. Nash equilibrium of the game $g_1$ is $({_{1}C},{_{1}D})$. Pair of rational action for player $2$ is  $({_{1}C},{_{1}D})_{2}$. Given the dominant strategy $_{1}S^{1}_{2}$ for player $2$ produces responses of player $1$ to this strategy pairs of rational actions. So pairs of rational actions for player $1$ is $({_{1}D},{_{1}D})_{1}$ and $({_{1}C},{_{1}D})_{1}$. Based on players being rational, player $2$ chooses dominant strategy $_{1}S^{1}_{2}$ and player $1$ chooses pair of rational actions  $({_{1}C},{_{1}D})_{1}$ to continue the game. As player $2$ wish to maintain his superiority, strategy $_{1}S^{1}_{2}$ end to Bluff game and pair of rational actions $({_{1}C},{_{1}D})_{1}$ ends to Mixed Harmony game. As a result, game $g_1$ is producer of Mixed Harmony game $g_2$ and Bluff game $g_3$.

Game $g_2$ is a strategy maker of order $(2, 2)$. In this game dominant strategy $_{2}S^{1}_{i}$ for player $i$ is cooperation and dominated strategy $_{2}S^{2}_{i}$ is defect. The game Nash equilibrium is $({_{2}C},{_{2}C})$. Pareto dominant Nash equilibrium compared to pair of actions is $({_{2}D},{_{2}D})$. As a result, the only pair of rational actions for both players is $({_{2}C},{_{2}C})_{1,2}$. Based on players being rational, the player $1$ chooses dominant strategy $_{2}S^{1}_{1}$ and player $2$ chooses pair of rational actions $({_{2}C},{_{2}C})_{2}$ to continue the game. As player $1$ desires cooperation, the dominant strategy $_{2}S^{1}_{1}$ ends to Pure Harmony game $g_4$. Pair of rational actions $({_{2}C},{_{2}C})_{2}$ ends to Stag Hunt $g_5$. Consequently, game $g_2$ is producer of games $g_4$ and $g_5$.

Game $g_3$ is strategy maker of order $(2, 1)$. In this game dominant  strategy $_{3}S^{1}_{2}$ for player $2$ is defect and dominated  strategy $_{3}S^{2}_{2}$ is cooperation. Game $g_3$ for player $1$ isn't strategy maker. Nash equilibrium of game $g_3$ is $({_{3}C},{_{3}D})$. Pair of rational actions for player 2 is $({_{3}C},{_{3}D})_{2}$. Considering dominant strategy $_{3}S^{1}_{2}$ player $2$, produces responses of player $1$ to this strategy pair of rational actions. So, pairs of rational actions for player $1$ is $(D_3,{_{3}D})_{1}$ and $({_{3}C},{_{3}D})_{1}$. Based on rationality of players, and also player $2$ desires to maintain his superiority chooses dominant strategy $_{3}S^{1}_{2}$ to continue the game. Dominant strategy $_{3}S^{1}_{2}$ ends to game $g_5$. As player $1$ desires cooperation in every stage doesn't choose a motion from this game to continuation.

Game $g_4$ is strategy maker of order $(2, 2)$. In this game dominant strategy $_{4}S^{1}_{i}$ for player $i$ is cooperation and his dominated strategy $_{4}S^{2}_{i}$ is defect. Pair of action $({_{4}C},{_{4}C})$ is Pareto dominant compared to pair of actions $({_{4}D},{_{4}D})$. Nash equilibrium of game and the only pair of rational action for both players is $({_{4}C},{_{4}C})_{1,2}$.

Game $g_5$ produced by dominant strategy $_{3}S^{1}_{2}$ of game $g_3$ and pair of rational actions $({_{2}C},{_{2}C})_{2}$ of game $g_2$. Therefore, it can be concluded that player $2$ for the game continuation has involved in dichotomy between choosing dominant strategy and pair of rational action that the dichotomy results in choosing of game $g_5$ for play continuation. Game $g_5$ is strategy maker of order $(2,0)$. Pair of actions $(C_5,C_5)$ is dominant Pareto compared to pair of actions $({_{5}D},{_{5}D})$. Nash equilibrium of game and players' pairs of rational actions is  $({_{5}C},{_{5}C})_{1,2}$ and $({_{5}D},{_{5}D})_{1,2}$.

Based on players being rational, they conclude from game $g_4$ and $g_5$ to cooperate with each other. Hence, in game $g_4$ pair of rational action $({_{4}C},{_{4}C})_{1,2}$ and in game $g_5$, they choose pair of rational action of dominant Pareto $({_{5}C},{_{5}C})_{1,2}$. As a result, these two games are producers of the No Conflict game $g_6$. In this step, the players have no appetence to continue. In the following, we obtain strategies and pairs of rational actions in Fig. \ref{fig:Peace}, for game $g_1$, we have:
\begin{align*}
&\phi_1(g_1)={_{1}\mathcal{S}_1}= \emptyset\\
& \phi_2(g_1)={_{1}\mathcal{S}_2}=\{{_{1}S^{1}_{2}}, {_{1}S^{2}_{2}}\}\\
&\phi^{'}_{1}(g_1)={_{1}\mathcal{A}^{'}_{1}}=\{({_{1}C},{_{1}D)_1}, {(_{1}D},{_{1}D)_1}\}=\{(2,4), (1,3)\}\\
&\phi^{'}_{2}(g_1)={_{1}\mathcal{A}^{'}_{2}}=\{({_{1}C}, {_{1}D})_2\}=\{(2,4)\}
\end{align*}

For game $g_2$, we have:
\begin{align*}
&\phi_1(g_2)={_{2}\mathcal{S}_1}=\{{_{2}S^{1}_{1}}, {_{2}S^{2}_{1}}\}\\
& \phi_2(g_2)={_{2}\mathcal{S}_2}=\{{_{2}S^{1}_{2}}, {_{2}S^{2}_{2}}\}\\
&\phi^{'}_{1}(g_2)={_{2}\mathcal{A}^{'}_{1}}=\{ {(_{2}C},{_{2}C)_1}\}=\{(4,4)\}\\
&\phi^{'}_{2}(g_2)={_{2}\mathcal{A}^{'}_{2}}=\{ (_{2}C,{_{2}C})_2\}=\{(4,4)\}
\end{align*}

For game $g_5$ we have:
\begin{align*}
&\phi_1(g_5)={_{5}\mathcal{S}_1}=\emptyset\\
& \phi_2(g_5)={_{5}\mathcal{S}_2}=\emptyset\\
&\phi^{'}_{1}(g_5)={_{5}\mathcal{A}^{'}_{1}}=\{({_{5}C},{_{5}C})_1, {(_{5}D},{_{5}D)_1}\}=\{(4,4), (2,2)\}\\
&\phi^{'}_{2}(g_5)={_{5}\mathcal{A}^{'}_{2}}=\{({_{5}C}, {_{5}C})_2, (_{5}D,{_{5}D})_2\}=\{(4,4), (2,2)\}
\end{align*}

Functions of the game move are as follows:
\begin{align*}
&\varphi_{2}(m^{1}_{1})=\varphi_{2}({_{1}S^{1}_{2}})=g_1g_3\\
&\varphi_{1}(m^{1}_{1})=\varphi_{1}({_{2}S^{1}_{1}})=g_2g_4\\
&\varphi_{1,2}(m^{2}_{1})=\varphi_{1,2}(({_{5}C}, {_{5}C})_{1,2})=g_5g_6
\end{align*}

 The system history is as follows:
 \begin{align*}
&H=\Big\{\emptyset, \big\{g_1, {_{1}S^{1}_{2}}, ({_{1}C},{_{1}D})_{1}\big\},  \big\{({_{1}C},{_{1}D})_{1},
\{g_2, {_{2}S^{1}_{1}}, ({_{2}C},{_{2}D})_{1, 2}\}\big\}, \big\{{_{1}S^{1}_{2}},\{g_3, {_{3}S^{1}_{2}}\}\big\},\\
&\;\;\;\;\;\;\;\;\;\;\;\ \big\{{_{2}S^{1}_{1}},\{g_4, ({_{4}C},{_{4}C})_{1,2}\}\big\},\big\{{_{3}S^{1}_{2}}, ({_{2}C},{_{2}C})_{ 2}, \{g_5, ({_{5}C},{_{5}C})_{1,2}\}\big\},\\
 &\;\;\;\;\;\;\;\;\;\;\;\ \big\{({_{4}C},{_{4}C})_{1,2}, ({_{5}C},{_{5}C})_{1,2}, \{g_6\}\big\}\Big\}
\end{align*}

Dynamic system of the above games shows that what games players need to perform to reach a mutual and satisfactory agreement. In many negotiations between countries such systems as the above system can designed. From the above example, it is concluded that Nash equilibrium of a game is a pair of rational actions but pair of rational actions isn't necessarily Nash equilibrium.

%%%%%%%%%%%%%%%%%%%%%%%%%%%%%%%%%%%%%%%%%%%%%%%%%%%%%%%%%%%%%%%%%%%%%%%%%%%%%%%%%%%%%%%%%%%%%%%%%%%%%%%%%%%%%%%%%%%%%%%%%%%%%%%%%%%%%%%%%%%%
%%%%%%%%%%%%%%%%%%%%%%%%%%%%%%%%%%%%%%%%%%%%%%%%%%%%%%%%%%%%%%%%%%%%%%%%%%%%%%%%%%%%%%%%%%%%%%%%%%%%%%%%%%%%%%%%%%%%%%%%%%%%%%%%%%%%%%%%%%%%%
\section{Second Persian Gulf War}\label{S.3}

America entered in Iraq in the shadow of slogans such as struggle with terrorism, world peace, granting democracy and freedom. Here, using the dynamic system of games, we will model the clash between the United States and Iran at the time of American troop presence in Iraq. To this end, we divide this time interval into five periods. And in each period, we will examine static games with complete information that has occurred. The first period before the start of the attack that is shown in the form of game $ g_1 $. The second period from March 20, 2003 until late 2004, when each player, based on his forces and facilities, was trying to fulfill his goals that includes games $ g_2 $ and $ g_3 $. The third period is from the late 2004 to the end of the Bush administration, which includes games $ g_4 $ and  $ g_5 $. The fourth period with the arrival of the new administration in America until the end of 2010, that including game $ G_6 $. Eventually, on December 15, 2011, with the total withdrawal of US forces from Iraq, the system stops which includes game $ g_7 $. For more information about Second War of Persian Gulf, refer to references \cite{ehteshami2003iran, fawn2006iraq, kumar2006media}.

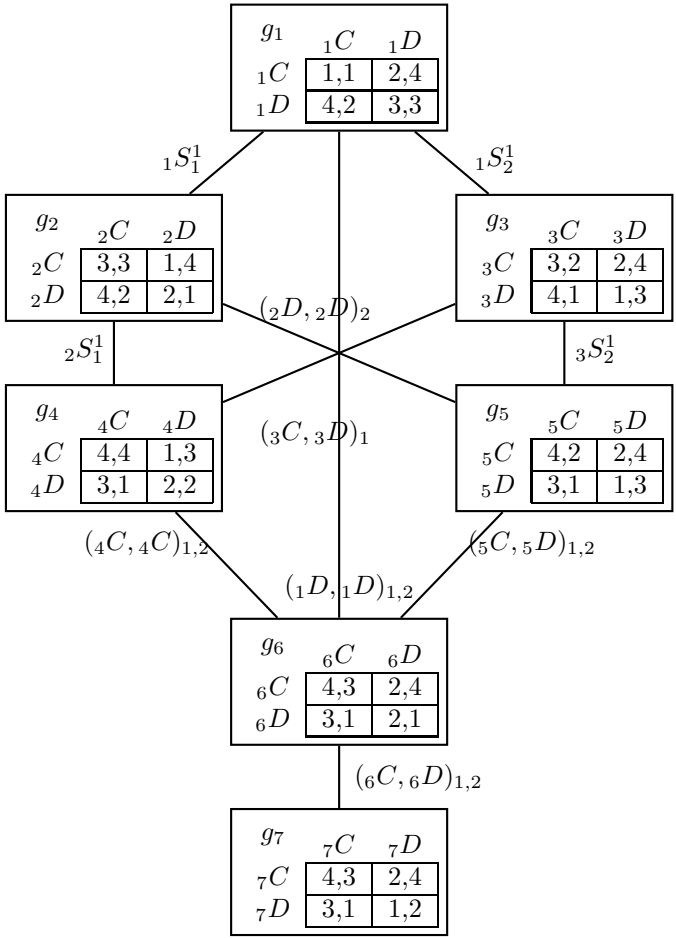
\begin{figure}
\centering
\begin{tikzpicture}
\node [ opmobject, xshift=-20] (1) {\label{tabel:label}\begin{tabular}{c|c|c|}
\multicolumn{1}{c}{$g_1$} & \multicolumn{1}{c}{} & \multicolumn{1}{c}{} \\[-2.5mm]
  \multicolumn{1}{c}{} & \multicolumn{1}{c}{${_{1}C}$} & \multicolumn{1}{c}{${_{1}D}$} \\ \cline{2-3}
   ${_{1}C}$ & 1,1 & 2,4 \\ \cline{2-3}
   ${_{1}D}$ & 4,2 & 3,3 \\ \cline{2-3}
\end{tabular}};
\node [  below=of 1, xshift=-59 pt, yshift=27 pt] (Q){$_{1}S^{1}_{1}$};
\node [ opmobject, below=of 1, xshift=-85 pt, yshift=5 pt] (2) {\begin{tabular}{c|c|c|}
\multicolumn{1}{c}{$g_2$} & \multicolumn{1}{c}{} & \multicolumn{1}{c}{} \\[-2.5mm]
  \multicolumn{1}{c}{} & \multicolumn{1}{c}{${_{2}C}$} & \multicolumn{1}{c}{${_{2}D}$} \\ \cline{2-3}
   ${_{2}C}$ & 3,3 & 1,4 \\ \cline{2-3}
   ${_{2}D}$ & 4,2 & 2,1 \\ \cline{2-3}
\end{tabular}};
\node [  below=of 1, xshift=59 pt, yshift=27 pt] (Q){$_{1}S^{1}_{2}$};
\node [ opmobject, below=of 1, xshift=85 pt, yshift=5 pt] (3) {\begin{tabular}{c|c|c|}
\multicolumn{1}{c}{$g_3$} & \multicolumn{1}{c}{} & \multicolumn{1}{c}{} \\[-2.5mm]
  \multicolumn{1}{c}{} & \multicolumn{1}{c}{${_{3}C}$} & \multicolumn{1}{c}{${_{3}D}$} \\ \cline{2-3}
   ${_{3}C}$ & 3,2 & 2,4 \\ \cline{2-3}
   ${_{3}D}$ & 4,1 & 1,3 \\ \cline{2-3}
\end{tabular}};
\node [  below=of 2, xshift=-11 pt, yshift=27 pt] (Q){$_{2}S^{1}_{1}$};
\node [ opmobject, below=of 2,  xshift=0 pt, yshift=5 pt] (4) {\begin{tabular}{c|c|c|}
 \multicolumn{1}{c}{$g_4$} & \multicolumn{1}{c}{} & \multicolumn{1}{c}{} \\[-2.5mm]
  \multicolumn{1}{c}{} & \multicolumn{1}{c}{${_{4}C}$} & \multicolumn{1}{c}{${_{4}D}$} \\ \cline{2-3}
   ${_{4}C}$ & 4,4 & 1,3 \\ \cline{2-3}
   ${_{4}D}$ & 3,1 & 2,2 \\ \cline{2-3}
\end{tabular}};
\node [  below=of 1, xshift=-9 pt, yshift=-30 pt] (Q){(${_{2}D},{_{2}D})_{2}$};
\node [  below=of 1, xshift=-9 pt, yshift=-77 pt] (Q){(${_{3}C},{_{3}D})_{1}$};
\node [  below=of 3, xshift=12 pt, yshift=27 pt] (Q){$_{3}S^{1}_{2}$};
\node [ opmobject, below=of 3,  xshift=0 pt, yshift=5 pt] (5) {\begin{tabular}{c|c|c|}
\multicolumn{1}{c}{$g_5$} & \multicolumn{1}{c}{} & \multicolumn{1}{c}{} \\[-2.5mm]
  \multicolumn{1}{c}{} & \multicolumn{1}{c}{${_{5}C}$} & \multicolumn{1}{c}{${_{5}D}$} \\ \cline{2-3}
   ${_{5}C}$ & 4,2 & 2,4 \\ \cline{2-3}
   ${_{5}D}$ & 3,1 & 1,3 \\ \cline{2-3}
\end{tabular}};
\node [  below=of 1, below=of 1, xshift=4 pt, yshift=-135 pt] (Q){ $({_{1}D},{_{1}D})_{1,2}$};
\node [  below=of 4, left=-40 pt, yshift=-12 pt] (Q){$({_{4}C},{_{4}C})_{1,2}$};
\node [  below=of 5, right=-40 pt, yshift=-12 pt] (Q){ $({_{5}C},{_{5}D})_{1,2}$};
\node [ opmobject, below=of 1,  xshift=0 pt, yshift=-155 pt] (6) {\begin{tabular}{c|c|c|}
\multicolumn{1}{c}{$g_6$} & \multicolumn{1}{c}{} & \multicolumn{1}{c}{} \\[-2.5mm]
  \multicolumn{1}{c}{} & \multicolumn{1}{c}{${_{6}C}$} & \multicolumn{1}{c}{${_{6}D}$} \\ \cline{2-3}
   ${_{6}C}$ & 4,3 & 2,4 \\ \cline{2-3}
   ${_{6}D}$ & 3,1 & 2,1 \\ \cline{2-3}
\end{tabular}};
\node [  below=of 6, xshift=30 pt, yshift=25 pt] (Q){(${_{6}C},{_{6}D})_{1,2}$};
\node [ opmobject, below=of 6,  xshift=0 pt, yshift=5 pt] (7) {\begin{tabular}{c|c|c|}
\multicolumn{1}{c}{$g_7$} & \multicolumn{1}{c}{} & \multicolumn{1}{c}{} \\[-2.5mm]
  \multicolumn{1}{c}{} & \multicolumn{1}{c}{${_{7}C}$} & \multicolumn{1}{c}{${_{7}D}$} \\ \cline{2-3}
   ${_{7}C}$ & 4,3 & 2,4 \\ \cline{2-3}
   ${_{7}D}$ & 3,1 & 1,2 \\ \cline{2-3}
\end{tabular}};
\path  (1) edge (2);
\path  (1) edge (3);
\path  (1) edge (6);
\path  (2) edge (4);
\path  (3) edge (4);
\path  (2) edge (5);
\path  (3) edge (5);
\path  (4) edge (6);
\path  (5) edge (6);
\path  (6) edge (7);
\end{tikzpicture}
\caption{Dynamic system of games between America and Iran}
\label{fig:Iraq War}
\end{figure}

Invasion of Iraq or Second Persian Gulf War resulted in production of many games among countries of Middle-East region and other countries including the game between two countries, for example, one can refer to Iran and America. We consider America as row player (player 1) and Iran as column player (player 2) in Fig. \ref{fig:Iraq War}. The game between these two countries at the beginning of war was Anti-Chicken game. In this game, America has two actions: either it wouldn't attack ${_{1}C}$ or attack to Iraq ${_{1}D} $. Iran was able to participate in the attack and make cooperation with America ${_{1}C}$ or wouldn't participate in the attack to Iraq and make defect ${_{1}D}$. In Anti-Chicken game, the two players have dominant strategy $_{1}S^{1}_{i}$, defection and dominated strategy $_{1}S^{2}_{i}$, cooperation. Nash equilibrium of game is $({_{1}D},{_{1}D})$.

These countries with different and conflicting purposes and objectives entered in the conflict arena. Clearly, every country attempts to realize its most and maximum objectives; but in the way of achieving goals of each player, there is other player’s objectives and possibilities. The more number and power of advocate forces of a country, the more its feasibility of purposes and objectives. America objectives in Iraq can be considered a part of Middle-East Strategy of this country. Therefore, America totally pursues a government in Iraq that firstly, doesn't prevent  keeping security and survival of Israel, secondly, keeping regular energy flow toward west and thirdly, the aim of prevention of establishment of a anti-American government and anti-western \cite{khalilzad1998sources}. Iran totally pursues a government in Iraq that firstly, America’s withdrawal from Iraq and its undermining in the case of insisting on presence in Iraq, secondly, prevention of establishment of a regime opposing Iran and thirdly, promotion of Shia groups position in Iraq scene.

Hence, players don't desire to cooperate and continue the game based on their strategic preferences. Dominant strategy for both players is $_{1}S^{1}_{i}$ and the pair of rational actions of game is $({_{1}D},{_{1}D})_{1,2}$. Based of players being rational, player $1$ chooses dominant strategy $_{1}S^{1}_{1}$ and player 2 chooses dominant strategy $_{1}S^{1}_{2}$ to continue.

Dominant strategy $_{1}S^{1}_{1}$ ends to Bluff game $g_2$. In $g_2$, America has two actions: either it takes control of Iraq with synergy of Iran, that is, it makes cooperation ${_{2}C}$ with Iran or take control of Iraq without synergy of Iran and wouldn't cooperate ${_{2}D}$ with Iran. In this game, Iran has two actions: either it makes cooperation ${_{2}C}$ with America to control Iraq or wouldn't make cooperation  ${_{2}D}$ with America. $g_2$ is strategy maker of order $(2, 1)$ and producer of dominant strategy $_{2}S^{1}_{1}$, defection  and dominated strategy $_{2}S^{2}_{1}$, cooperation. This game isn't strategy maker for player $2$. Players have pairs of rational actions $({_{2}D},{_{2}C})_{1,2}$ and $({_{2}D},{_{2}D})_{2}$. Based on players being rational, player $1$ chooses dominant strategy $_{2}S^{1}_{1}$ to continue and player $2$ chooses pair of rational actions $({_{2}D},{_{2}D})_{2}$ to continue.

America War in Iraq in the classic form lasting for three weeks. By Baghdad collapse and runaway of Saddam and his boys, Saddam's regime was ruined, but despite of American’s initial impression, this was a superficial victory. They entered in Iraq with an attritional war with rest of Baathist Regime forces and Sonni groups that gradually its intensity and scope was increased. Volunteer Sonni forces that mostly had associated with Alqaede, entered in Iraq from other countries.  Their attack against Iraq Shiites provoked an ethnical war in this country and made its situation more complicated \cite{davis2016presidential}.

Dominant strategy $_{1}S^{1}_{2}$ ends to Bully game $g_3$. In $g_3$, Iran has two actions: both makes attempts to reach its objectives and wouldn't make cooperation ${_{3}D}$  or withdraw from its objectives and cooperation with America ${_{3}C}$. Also, America has two actions: either want synergy of Iran to control Iraq ${_{3}C}$ or wouldn't choose cooperation of Iran ${_{3}D}$. $g_3$ is strategy maker of order $(2, 1)$ and producer of dominant strategy of defect $_{3}S^{1}_{2}$ and dominated strategy of cooperation $_{3}S^{2}_{2}$. This game for player $1$ isn't strategy maker. Players have pairs of rational actions $({_{3}C},{_{3}D})_{1,2}$ and $({_{3}D},{_{3}D})_{1}$. Based on players being rational, player $2$ chooses dominant strategy $_{3}S^{1}_{2}$ and player $1$ chooses pair of rational actions $({_{3}C},{_{3}D})_{1}$ to continue.

Iraq situation became more deteriorated daily, violence increased and hate of America was enhanced. America to prevent this situation attempted to reduce the violence level by holding elections and transferring power to Iraqis  \cite{baker2006iraq}. In the new strategy, to confront with threats, America persuaded increasing American forces in Iraq, increase and reinforcement of Iraqis forces capabilities to establish stability and peace, increasing participation of Sonni forces in power, pressure on Iran and Syria to reduce support of groups and preventing foreign forces into Iraq and attempt to negotiate with Iraq’s neighbor countries \cite{baker2006iraq}. By increasing American forces in Iraq, pressure on militant groups was added and America’ attempt to provide an appropriate base in order to participation of Iraqis tribes in government and the political process of this country and finally, negotiation with Iran and Syria to reduce support of competing groups within Iraq yielded and violence began to decline.

Game $g_4$ produced through dominant strategy $_{2}S^{1}_{1}$ from $g_2$ and pair of rational actions  $({_{3}C},{_{3}D})_{1}$ of $g_3$. Therefore, it can be concluded that player $1$ to continue his play has engaged in dichotomy between choosing dominant strategy and pair of rational actions that this dichotomy results in choosing Stag Hunt $g_4$ to continue. In $g_4$, America has two actions: either it takes control of Iraq, by adding forces and negotiation with Iran, that is, making cooperation with Iran ${_{4}C}$ or doesn't cooperate with Iran ${_{4}D}$ . Also, Iran also has two actions: either it makes cooperation to control Iraq ${_{4}C}$  or wouldn't make cooperation with America ${_{4}D}$. $g_4$ is strategy maker of order $(2, 0)$. Pair of actions  $({_{4}C},{_{4}C})$ is dominant Pareto compared to pair of actions $({_{4}D},{_{4}D})$. Nash equilibria and the game pairs of rational actions are $({_{4}C},{_{4}C})_{1,2}$ and $({_{4}D},{_{4}D})_{1,2}$.

Considering the power and forces had in Iraq, Iran, started to extort from America to reach its objectives. The Iraq National Parliament election was held in January 30, 2005 throughout the country. Most of Sonni groups boycotted it. Shiites Union and Kurds obtained 140 and 75 seats, respectively. Premiership of Iraq was devoted to Shiites, Presidency office to Kurds and Parliament administration to Sonni group \cite{katzman2006iraq}.

Game $g_5$ produced through dominant strategy $_{3}S^{1}_{2}$ from $g_3$ and pair of rational actions  $({_{2}D},{_{2}D})_{2}$ of $g_2$. In  Blackmailer game $g_5$, Iran has two actions: either intends to prevent establishment of a government opposing with itself and promotion of Shiite position and wouldn't make cooperation ${_{5}D}$ or make cooperation ${_{5}C}$ with America for establishment of a government aligned with America. Also, America also has two actions: either it supports establishment of a aligned government with Iran and makes cooperation to control violence in Iraq ${_{5}C}$ with Iran or oppose an aligned government with Iran and wouldn't make cooperation ${_{5}D}$. $g_5$ is strategy maker of order $(2,2)$. In this game, dominant strategy $_{5}S^{1}_{1}$ for player 1 is cooperation and dominated strategy $_{5}S^{2}_{1}$, is defection. Also for player $2$, the dominant strategy $_{5}S^{1}_{2}$ is defect and dominated strategy $_{5}S^{2}_{2}$ is cooperate. The game Nash equilibrium and the only pair of rational actions for both players is $({_{5}C},{_{5}D})_{1,2}$.

In Iraq, the process of changes in the final years of Bush and the first year of Obama showed that America's new strategy was effective in Iraq, violence has been controlled partly and tensions are abating. Continuation of this trend provides more ground for US forces withdrawal from Iraq. So, Obama government and America Congress have explicitly announced that America doesn't need establishing permanent military bases in Iraq. According to the reached agreements and policies announced by the American government, withdrawal of American military forces until August 2010 is realized and ended until end of year 2011. In December 15, 2011, America terminated officially its military presence in Iraq by holding an official ceremony in Baghdad airport located in Baghdad Green Region, in presence of America then Secretary of Defense.

According to players being rational, pair of rational actions $({_{1}D},{_{1}D})_{1,2}$, $({_{5}C},{_{5}D})_{1,2}$ and $({_{4}C},{_{4}C})_{1,2}$ chosen by players ends to Trickery game $g_6$. In other words, $g_4$ and $g_5$ are producer of $g_6$. In $g_6$, America has two actions: either it reduces its forces  ${_{6}C}$ and gradually withdraws from Iraq or increases its forces in Iraq ${_{6}D}$ and imports more military equipments into Iraq. Also, Iran has two actions: either it puts pressure on America in assigning control of Iraq to a new government ${_{6}D}$ or cooperates with America  ${_{6}C}$. The game Nash equilibria are $({_{6}C},{_{6}D})$ and $({_{6}D},{_{6}D})$. In $g_6$, weakly dominant strategy $_{6}S^{1}_{1}$ for player $1$ is reducing forces and weakly dominated strategy $_{6}S^{2}_{1}$ is increasing forces. Also, for player $2$ weakly dominant strategy $_{6}S^{1}_{2}$ is to put pressure on America and weakly dominated strategy $_{6}S^{2}_{2}$ is cooperation with America. The players' pairs of rational actions are $({_{6}C},{_{6}D})_{1,2}$, $({_{6}D},{_{6}D})_{1,2}$ and $({_{6}C},{_{6}C})_{2}$.

Based on rationality of players and strategic preferences, players selected pair of rational actions $({_{6}C},{_{6}D})_{1,2}$ that ends to Hegemony game $g_7$. In $g_7$, America has two actions: either withdraw its forces from Iraq ${_{7}C}$ or maintain its forces in Iraq ${_{7}D}$. Also, Iran has two actions: either cooperate with America ${_{7}C}$ or wouldn't cooperate with America ${_{7}D}$. The game Nash equilibrium is $({_{7}C},{_{7}D})$. In this step, the players have no appetence to continue. According to our definition, $g_7$ is a final node and the system is completed. Dynamic system of strategic games between Iran and America is represented by graphs in Fig. \ref{fig:Iraq War}. History of system is as follows:
 \begin{align*}
&H=\Big\{\emptyset, \big\{g_1, {_{1}S^{1}_{1}}, {_{1}S^{1}_{2}} \big\}, \big\{{_{1}S^{1}_{1}},\{g_2, {_{2}S^{1}_{1}}, ({_{2}D},{_{2}D})_{2}\}\big\}, \big\{{_{1}S^{1}_{2}},\{g_3, {_{3}S^{1}_{2}}, ({_{3}C},{_{3}D})_{1}\}\big\},~~ \\
 &\;\;\;\;\;\;\;\;\;\ \big\{{_{2}S^{1}_{1}}, ({_{3}C},{_{3}D})_{1}, \{g_4, ({_{4}C},{_{4}C})_{1, 2}\}\big\}, \big\{{_{3}S^{1}_{2}},({_{2}D},{_{2}D})_{2}, \{g_5, ({_{5}C},{_{5}D})_{1,2}\}\big\},\\
 &\;\;\;\;\;\;\;\;\;\ \big\{({_{4}C},{_{4}C})_{1,2}, ({_{5}C},{_{5}D})_{1,2}, ({_{1}D},{_{1}D})_{1,2}, \{g_6, ({_{6}C},{_{6}D})_{1,2}\}\big\},\\ &\;\;\;\;\;\;\;\;\;\ \big\{({_{6}C},{_{6}D})_{1,2}, \{g_7\}\big\}\Big\}.
\end{align*}

The above modeling shows that complete withdrawal of American forces from Iraq lead to more influential of Iran in the region. The Nash equilibrium $({_{6}D},{_{6}D})$ in game $g_6$ shows if America in this game prefer tactical preferences over strategic preferences, obtain better result and conditions to continue this system was changed.

%%%%%%%%%%%%%%%%%%%%%%%%%%%%%%%%%%%%%%%%%%%%%%%%%%%%%%%%%%%%%%%%%%%%%%%%%%%%%%%%%%%%%%%%%%%%%%%%%%%%%%%%%%%%%%%%%%%%%%%%%%%%%%%%%%%%%
%%%%%%%%%%%%%%%%%%%%%%%%%%%%%%%%%%%%%%%%%%%%%%%%%%%%%%%%%%%%%%%%%%%%%%%%%%%%%%%%%%%%%%%%%%%%%%%%%%%%%%%%%%%%%%%%%%%%%%%%%%%%%%%%%%%%%
\section{Trickery game}\label{S.4}
We introduced a new game $g_6$ in Fig. \ref{fig:Iraq War}, called Trickery game. The trickery game is a asymmetric game $2\times2$ that examines difficult conditions of decision making between players. Consider two companies which compete for achieving a common source. Two players seek for more benefit from common source. Therefore, players request for a portion of source considering their potential and capability that determines bargaining power of players. One of two players has fewer capability to use from common source. Each player can choose or cooperation or defect to resolve conflict. So, set of actions of the players includes cooperation $C$ and defect $D$. We consider row player as player $1$ and column player as player $2$. player $1$ has fewer capability to use from common source. The trickery game table is given in Fig. \ref{fig:RS}.

\begin{figure}
\centering
\begin{tikzpicture}
\node [ opmobject] (4) {\begin{tabular}{c|c|c|}
\multicolumn{1}{c}{$G$} & \multicolumn{1}{c}{} & \multicolumn{1}{c}{} \\[-2.5mm]
\multicolumn{1}{c}{} & \multicolumn{1}{c}{C} & \multicolumn{1}{c}{D} \\ \cline{2-3}
C & 4,3 & 2,4 \\ \cline{2-3}
D & 3,1 & 2,1 \\ \cline{2-3}
\end{tabular}};
\end{tikzpicture}
\caption{ Payoff table for the Trickery game }
\label{fig:RS}
\end{figure}

The game has two Nash equilibrium $(C, D)$ and $(D, D)$. We call this game, Trickery game; because player $1$ can choose cooperation with player $2$ until the last moment and finally changes his action to defect by trickery. While, player 1 has weakly dominant action of cooperation, but assuming the choice of non-cooperation by player 2, player 1 can with cunning change his action that Reduces the payoff of player 2. This game is strategy maker of order $(2, 0)$. It has three pairs of rational actions $(C, C)$, $(D, D)$ and $(C, D)$.

%%%%%%%%%%%%%%%%%%%%%%%%%%%%%%%%%%%%%%%%%%%%%%%%%%%%%%%%%%%%%%%%%%%%%%%%%%%%%%%%%%%%%%%%%%%%%%%%%%%%%%%%%%%%%%%%%%%%%%%%%%%%%%%%%%%%%%%%%%%%%%%%%%%%
%%%%%%%%%%%%%%%%%%%%%%%%%%%%%%%%%%%%%%%%%%%%%%%%%%%%%%%%%%%%%%%%%%%%%%%%%%%%%%%%%%%%%%%%%%%%%%%%%%%%%%%%%%%%%%%%%%%%%%%%%%%%%%%%%%%%%%%%%%%%%%%%%%%%

\section{Discussion}
The majority of results in game theory concern simple games with a few players and a few possible actions, characterizing them in terms of their equilibria. Game theory in static state use one game, but dynamic system of strategic games use several games to model an event. With emphasis on players’ rationality, we present new properties of strategic games, which result in the dynamics existing in interactions among players. Since the classic theory of games lacks an explicit treatment of the dynamics of rationadeliberation, dynamic system of strategic games can be seen, as filling an important lacuna of classic game theory.

In this study, with a new attitude toward $2\times 2$ games that Nash equilibrium is one of the most important properties of these games, we achieved new properties such as strategies generated by a game and pairs of rational actions. According to this feature, strategic games was divided into two classes, strategy maker games and games that aren't strategy maker. Strategy maker games itself are of two groups: either strategy maker of order $(2, 2)$, that is, the game is strategy maker for both players or strategy maker of order $(2,1)$, that is, game just is strategy maker for one player and not for other one. Also, property of pair of rational actions for a game was raised. A $2\times 2$ game based on payoff of players has one, two, three or four pairs of rational actions or doesn't have. Also, games that aren't strategy maker itself are of two groups: either has pair of rational actions or does'nt have. The $2\times 2$ games can be classified based on number of pair of rational actions as well.

With  emphasis on players' rationality and with  help of strategy maker being property of a game and pairs of rational actions of a game, how new games are produced was discussed through a game. This choice that is taken place based on player's rationality is called the player's strategic preferences. Player’s action selection in the game conditions is called tactical preferences. Depending on present and future conditions, a player may prefer strategic preferences on tactical preferences and vice versa. Also, this attitude enables players to find in negotiations and conflicts a solution to reach a mutual agreement. Therefore, to reach this agreement, we outlined dynamic system of games with strategic games. In this system players according their rationality determine the path of achieving agreement. We use graph tools to display this system. In each node of the graph, there is a strategic $2\times 2$  game and both players are able to decide. Moves or edges of this graph are generated strategies or pairs of rational actions. The system history includes games and moves chosen by players. Each player selects a move which has more benefit for him according to his rationality. These stages continue by moving from initial node until players reach to agreement in one node that this node is called final node of system. By cutting periods and environmental conditions, dynamic system of games can applies for modeling past events. Also, this system can be used to model present and future conditions. With this system, relations between players (countries, companies, humans,…) can be modeled from the first relationship to the last one. For example, e will examine the clash between the United States and Iran from the time of start of the invasion of Iraq until the complete withdrawal of U.S. troops using dynamic system of strategic games. The result show that, whit withdrawal of American forces from Iraq results in Iran hegemony in the region. Another application of this system is to relate  games and examine their impact on each other. Moreover, we introduce an asymmetric game $2\times2$ which one of players can with cunning change his action that reduces the payoff other player.

Dynamic system of games says that in most of the cases, player must wait for other players' move. The more limitations a game theory use, the more unlikely its prediction for future is and the less limitations we need to heavy reasoning and its analysis would be. We believe that maybe complete modeling of an event through one game wouldn't be possible but there is more chance with several games.

%%%%%%%%%%%%%%%%%%%%%%%%%%%%%%%%%%%%%%%%%%%%%%%%%%%%%%%%%%%%%%%%%%%%%%%%%%%%%%%%%%%%%%%%%%%%%%%%%%%%%%%%%%%%%%%%%%%%%%%%%%%%%%%%%%%%%%%%%%%%%%%%%%%%%%%     %%%%%%%%%%%%%%%%%%%%%%%%%%%%%%%%%%%%%%%%%%%%%%%%%%%%%%%%%%%%%%%%%%%%%%%%%%%%%%%%%%%%%%%%%%%%%%%%%%%%%%%%%%%%%%%%%%%%%%%%%%%%%%%%%%%%%%%%%%%%%%%%%%%%%%%

\end{document}